\documentclass[11pt]{article}
\usepackage{latexsym}
\usepackage{graphicx}
\usepackage{caption2}
\usepackage{psfrag}
\newcommand{\be}{\begin{equation}}
\newcommand{\ee}{\end{equation}}
\newcommand{\bea}{\begin{eqnarray}}
\newcommand{\eea}{\end{eqnarray}}

\newcommand{\lesssim}{ {\
\lower-1.2pt\vbox{\hbox{\rlap{$<$}\lower5pt\vbox{\hbox{$\sim$}}}}\
} }
\newcommand{\gtrsim}{ {\
\lower-1.2pt\vbox{\hbox{\rlap{$>$}\lower5pt\vbox{\hbox{$\sim$}}}}\
} }

\input epsf

\begin{document}

\begin{titlepage}

\begin{flushright}
\end{flushright}
\vspace*{1.5cm}
\begin{center}
{\Large \bf Long-distance dimension-eight operators in $B_K$}\\[3.0cm]

{\bf O. Cat{\`a}} and {\bf S. Peris}\\[1cm]

 Grup de F{\'\i}sica Te{\`o}rica and IFAE\\ Universitat
Aut{\`o}noma de Barcelona, 08193 Barcelona, Spain.\\[0.5cm]

\end{center}

\vspace*{1.0cm}

\begin{abstract}

Besides their appearance at short distances  $\gtrsim 1/M_W$, local dimension-eight
operators also contribute to kaon matrix elements at long distances of order $\gtrsim
1/\mu_{ope}$, where $\mu_{ope}$ is the scale controlling the Operator Product Expansion
in pure QCD, without weak interactions. This comes about in the matching condition
between the effective quark Lagrangian and the Chiral Lagrangian of mesons. Working in
dimensional regularization and in a framework where these effects can be systematically
studied, we calculate the correction from these long-distance dimension-eight operators
to the renormalization group invariant $\hat{B}_K$ factor of $K^0-\bar{K}^0$ mixing, to
next-to-leading order in the $1/N_c$ expansion and in the chiral limit. The correction is
controlled by the matrix element $\langle0|\bar{s}_L\tilde{G}_{\mu\nu}\gamma^{\mu}d_L
|K^0\rangle$, is small, and lowers $\hat{B}_K$.

\end{abstract}

\end{titlepage}
\section{Introduction}

In order to study the physics at the scale of the K meson one can integrate out the
fields describing all the other heavier particles. In the Standard Model, this
integration generates a local $\Delta S=2$ operator of the form\cite{Buras}
\begin{equation} \label{Heff}
{\cal{H}}_{\Delta
S=2}^{eff}=\frac{G_F^2m_W^2}{4\pi^2}[\lambda_c^2F_1+\lambda_t^2F_2+2\lambda_c\lambda_tF_3]
\ C_{\Delta  S=2}(\mu){\cal{Q}}_{\Delta S=2}(x)
\end{equation}
where $C_{\Delta S=2}$ is the Wilson coefficient given (in the large-$N_c$ limit) by
\begin{equation}\label{wilson}
    C_{\Delta S=2}(\mu)=\left[1+
\frac{\alpha_s(\mu)}{\pi}\left(\frac{1433}{1936}+\frac{\kappa}{8}\right)\right]
\left(\frac{1}{\alpha_s(\mu)}\right)^{\frac{9}{11 N_c}} \ .
\end{equation}
The Wilson coefficient encodes the ultraviolet physics which has been integrated out and
is computed in perturbation theory. Consequently, it depends on the renormalization scale
($\mu$) and scheme ($\kappa$). In Eq.(\ref{wilson}) $\kappa=0 \ \mathrm{or} -4$ depending
on whether one has used naive dimensional regularization or the 't Hooft-Veltman scheme
(respectively) in the calculation. The operator ${\cal{Q}}_{\Delta S=2}(x)$ in
Eq.\,(\ref{Heff}) reads
\begin{equation} \label{fourquark}
{\cal{Q}}_{\Delta
S=2}(x)=\bar{s}_L(x)\gamma_{\mu}d_L(x)\bar{s}_L(x)\gamma^{\mu}d_L(x)\quad , \quad
\psi_{R,L}\equiv \frac{1\pm \gamma_{5}}{2}\ \psi\ .
\end{equation}
 The $F_i$ are functions of the heavy particles which have been integrated out and their
expression can be found in \cite{Buras}.

The invariant bag factor $\hat{B}_K$ is then defined in terms of this effective weak
hamiltonian as
\begin{equation} \label{definition}
\left<\bar{K}^0|C_{\Delta  S=2}(\mu){\cal{Q}}_{\Delta
      S=2}(0)|K^0\right>\equiv\,\frac{4}{3}F_K^2 M_K^2\hat{B}_K \
\end{equation}
and, by construction, it is renormalization scheme and scale independent.

Because of confinement and chiral symmetry breaking, there is another effective
description which involves directly the lowest-lying mesons as degrees of freedom in a
chiral Lagrangian organized in powers of momenta and meson masses. In the case of $\Delta
S=2$
 interactions, the lowest-order operator in this chiral Lagrangian is given by

\begin{equation} \label{bosonization}
{\cal{L}}^{\chi}_{\Delta S=2}=\frac{F_0^4}{4}g_{\Delta
S=2}(\mu)\, \mathrm{Tr}\left(\lambda_{32}U^{\dagger}r^{\mu}U\lambda_{32}
U^{\dagger}r_{\mu}U\right)\ ,
\end{equation}
where $(\lambda_{32})_{ij}=\delta_{i3}\delta_{2j}$ is a (spurion) matrix in flavor space,
$F_0 \simeq 0.087$ GeV is the pion decay constant in the chiral limit and $U$ is a
$3\times 3$ unitary matrix collecting the Goldstone boson excitations and transforming as
$U\rightarrow R U L^{\dag}$ under a combined unitary flavor rotation $(R,L)$ of
$SU(3)_{R} \times SU(3)_L$. In (\ref{bosonization}) $g_{\Delta S=2}(\mu)$ is a low energy
coupling to be determined, as usual, by a matching condition to the Lagrangian
(\ref{Heff}). This condition is a crucial ingredient in the construction of the chiral
Lagrangian (\ref{bosonization})and secures that the ultraviolet behavior of
(\ref{bosonization}) be the same as that of (\ref{fourquark}) so that physics is
independent of conventions such as the renormalization scale $\mu$, the renormalization
scheme (i.e. NDR vs HV), etc...

  Although matching conditions are commonplace in the construction of
effective Lagrangians such as (\ref{Heff}) in a perturbative context, it is only recently
that their crucial role for the construction of the Lagrangian (\ref{bosonization}) has
been appreciated also in the nonperturbative context of analytic calculations of
electroweak matrix elements of light mesons\cite{Tempe}.

In the particular case of $B_K$ and $g_{\Delta  S=2}(\mu)$, a convenient Green's function
to do the matching with is\cite{PdeR}
\begin{equation} \label{Green}
{\cal{W}}_{\mu\alpha\nu\beta}^{LRLR}(q,l)=\lim_{l\rightarrow 0}\,i^3
\int\,d^4x\ d^4y\ d^4z\,e^{iq·x}e^{il·(y-z)}
<0|T\{L^{\bar{s}d}_{\mu}(x)R^{\bar{d}s}_{\alpha}(y)L^{\bar{s}d}_{\nu}(0)
R^{\bar{d}s}_{\beta}(z)\}|0>
\end{equation}
where
\begin{equation}
L^{\bar{s}d}_{\mu}(x)=\bar{s}(x)\gamma_{\mu}\frac{1-\gamma_5}{2}d(x)\qquad
R^{\bar{d}s}_{\mu}(x)=\bar{d}(x)\gamma_{\mu}\frac{1+\gamma_5}{2}s(x)
\end{equation}
are QCD chiral currents. It is a general property that chiral coupling constants such as
 $g_{\Delta S=2}$ in Eq. (\ref{bosonization}) can always be defined by means of QCD
 Green's functions --in the forward
limit-- which are chiral order parameters, and hence vanish to all orders in perturbation
theory in the chiral limit.

Performing a standard analysis of Ward identities leads to the following relation
\begin{equation}\label{ward}
\int\,d\Omega_q\ g^{\mu\nu}\
{\cal{W}}_{\mu\alpha\nu\beta}^{LRLR}(q,l)_{unfactorized}=
  \left(\frac{l^{\alpha}l^{\beta}}{l^2}-g^{\alpha\beta}\right){\cal{W}}^{LRLR}(Q^2)\quad
  ,\quad Q^2\equiv -q^2 \ ,
\end{equation}
where $\int d\Omega_q$ stands for the average over the momentum $q$ in $D$ dimensions,
namely
\begin{equation}\label{average}
    \int d\Omega_q \ q_{\mu} q_{\nu} \ f(q^2)= \frac{q^2 g_{\mu\nu}}{D}\  f(q^2)\ ,
\end{equation}
for a given function $f(q^2)$. Using then the definitions
\begin{equation}\label{def}
    z=\frac{Q^2}{\mu_{had}^2}  \qquad ,\qquad  {\cal{W}}^{LRLR}(z) =
    \frac{F_0^2}{z\mu_{had}^2}W(z)\ ,
\end{equation}
one can express $g_{\Delta S=2}$ in Eq. (\ref{bosonization})
as\cite{PdeR}
\begin{equation} \label{g}
g_{\Delta S=2}(\mu,\epsilon)=1-\frac{\mu_{had}^2}{32\pi^2F_0^2}
\left(\frac{4\pi\mu^2}{\mu_{had}^2}\right)^{\frac{\epsilon}{2}}
\frac{1}{\Gamma(2-\frac{\epsilon}{2})}\int_0^{\infty}dz\,z^{-\frac{\epsilon}{2}}W(z)\
.
\end{equation}
This equation expresses the matching condition between the chiral meson Lagrangian
(\ref{bosonization}) and the quark Lagrangian (\ref{Heff}). We emphasize that the scale
$\mu_{had}^2$ in Eq. (\ref{g}) is used solely for a rescaling of the momentum $Q^2$ and,
in principle, is totally arbitrary. In practice, however, there is always a residual
dependence on $\mu_{had}^2$ in the final result since calculations are done to a finite
order in, e.g., $\alpha_{s}$. Therefore this scale $\mu_{had}^2$ has to be numerically
large enough so as to make meaningful the truncated series in $\alpha_{s}$. Physically,
one can think of $\mu_{had}^2$ as the scale at which resonances are integrated out and
the Lagrangian (\ref{bosonization}) sets in.

>From Eq. (\ref{g}) it is clear that $g_{\Delta S=2}(\mu)$ requires the knowledge of the
Green's function $W(z)$ over the full range of momenta and regretfully, we hasten to say,
this information is not available. What is available, however, is its high- and low-$z$
behavior since they are given by the Operator Product and Chiral expansions,
respectively. Also, it is known that the analytic structure of $W(z)$ simplifies notably
in the large-$N_c$ limit becoming a meromorphic function, i.e. consisting of only poles,
with no cuts. However, lacking the solution to QCD at large $N_c$, the problem is still
too difficult to tackle even in this limit: the masses and residues of the infinite
number of resonances contributing to $W(z)$ are not known. Obviously, a further
approximation is necessary and this has been termed ``The Hadronic Approximation to
large-$N_c$ QCD'' (HA).

In order to motivate this approximation, it is important to realize that it is not the
point-wise behavior of $W(z)$ that is necessary, but only its integral. Moreover, $W(z)$
is defined fully in the euclidean regime. Therefore, it is reasonable to expect that a
smooth interpolation between the low-$z$ and the high-$z$ regimes of $W(z)$ may do a good
job in approximating the integral. Consequently we shall construct the HA as a rational
approximation\footnote{i.e. it can be written as a ratio of two polynomials.} to
large-$N_c$ QCD by keeping only a \emph{finite} number of zero-width resonances whose
masses and residues are obtained from matching to the first few coefficients of the
Operator Product and Chiral expansions. In the case of the coupling $g_{\Delta S=2}$ in
Eq. (\ref{g}), these expansions correspond to the high- and low-$z$ behavior of the
function $W(z)$, respectively.

The rational approximant so constructed, $W_{HA}(z)$, is therefore an interpolating
function between the high-$z$ and low-$z$ tails of $W(z)$ which it matches by
construction. The more terms one knows in the Operator Product and Chiral expansions, the
more resonances one can determine from the matching to construct the interpolator
$W_{HA}(z)$. Therefore the approximation is, in principle, improvable. In practice,
however, often only one or two terms in each expansion are known, so that one actually
has to borrow the masses of the resonances from the Particle Data tables and leave only
the residues to be determined by these matching conditions\footnote{If enough conditions
were known one could also determine the masses, of course. For a model in which this is
done, see Ref. \cite{GPdeR}}. There is a minimal requirement, though, which is that
$W_{HA}(z)$ \emph{has} to reproduce at least the leading non-vanishing term in the OPE.
This is indispensable for accomplishing cancelation of scale and renormalization scheme
dependence with Wilson coefficients, such as $C_{\Delta S=2}(\mu)$.
 This was done in the calculation of $\hat{B}_K$ in Ref.\,\cite{PdeR}. Several
 other observables have also been computed within this
framework\cite{theworks}\cite{MIT}.

We may now take the example of the matching condition (\ref{g}) to discuss the appearance
of local dimension-eight operators at relatively long distances. Indeed, the large-$z$
fall-off of the function $W(z)$ in this matching condition  between the chiral meson
Lagrangian and the quark Lagrangian is given by the OPE of the Green's function
(\ref{Green}). Dimension-eight operators (and higher) contribute to this OPE and thus
they appear in kaon matrix elements at rather long distances, i.e. \emph{without}
short-distance suppression factors such as  $1/M_W^2$. Notice that, despite being local,
these dimension-eight operators do not appear in the quark effective
Lagrangian\footnote{In a way, they correspond to having integrated all the meson
resonances out of the Lagrangian. This is why only Goldstone degrees of freedom appear in
(\ref{bosonization}). This integration can not be encoded in a quark effective operator
in the Lagrangian because this would give rise to a double counting problem: this quark
effective operator would generate both Goldstone mesons \emph{and} the resonances which
one thought one had integrated out!.}.

The appearance of these operators has been correctly emphasized in \cite{CDG},
particularly in approaches with the use of a momentum cutoff. However, we remark that
this contribution from dimension-eight operators is not caused by the cutoff and, in
fact, is also true even if one uses a purely dimensional regularization, as in
Eq.\,(\ref{g}). Clearly, a physical effect can not depend on the particular
regularization employed. What happens in approaches using a momentum cutoff $\Lambda$
(and does not happen in dimensional regularization) is that, by pure dimensional
analysis, local dimension-eight operators can also appear at distances $\gtrsim
1/\Lambda$ suppressed by $1/\Lambda^2$. \footnote{In the lattice jargon they are known as
$\mathcal{O}(a)$ effects. For us, $a\sim 1/\Lambda$.} Moreover, the authors of \cite{CDG}
have cautioned that these long-distance contributions from dimension-eight operators may
give potentially large contributions to weak matrix elements.

In our large-$N_c$ expansion approach, based on dimensional regularization, we shall see
that dimension-eight operators produce contributions which can be roughly characterized
as of $\mathcal{O}(\alpha_s \ \delta^{2}/\mu_{ope}^2)$, where $\delta^{2}$ is a
Goldstone-to-vacuum matrix element and $\mu_{ope}$ is the scale controlling the Operator
Product Expansion of correlators in pure QCD, i.e. \emph{without} weak interactions. To
be a bit more precise about this scale, we may define it in general as the momentum scale
above which the OPE in pure QCD starts making sense, even if this is only in the limited
sense of an asymptotic expansion. Phenomenologically this scale $\mu_{ope}$ may in
general be somewhat dependent on the quantum numbers involved, but it is usually of the
order of a resonance mass (i.e. $\sim 1$ GeV).

Besides the long-distance contributions alluded to above, there are other short-distance
contributions which are also due to local dimension-eight operators but which appear
after the integration of the $W$ boson or, in general, of any heavy field. These
contributions, unlike the former, \emph{are} encoded in the quark effective Lagrangian.
Indeed, when writing the effective Hamiltonian of Eq. (\ref{Heff})  one is neglecting
dimension-eight operators at short distances which appear below the charm mass, $m_c$,
after integration of this quark. Consequently, these dimension-eight operators appear
suppressed by the short distance scale $1/m_c^2$.  The size of this $1/m_c^2$ effect is
also an interesting problem in itself but, since it can be dealt with by ordinary
methods, it will not be the subject of the present work\footnote{Some $1/m_c$ effects
have been already considered, for example, in Ref. \cite{mc}.}. We plan to come back to
it in the future.

In this work we would like to quantitatively investigate the impact of the
$1/\mu_{ope}^2$-suppressed dimension-eight operators in approaches based on dimensional
regularization, such as the HA. If dimension-eight operators are to give an important
contribution in this case, they should correct significantly the lowest order term in the
OPE at large $Q^2$, i.e. the high-$z$ tail of functions such as $W(z)$ in the
determination of $g_{\Delta S=2}(\mu)$ in Eq. (\ref{g}). In the following we present a
new evaluation of $\hat{B}_K$ using the Hadronic Approximation to large-$N_c$ QCD with
inclusion of dimension-eight operators in the OPE. We now turn to the details of this
analysis.

After chiral corrections, the result in the $N_c\rightarrow \infty$ limit stems from the
full factorization of the operator (\ref{fourquark}) into two left-handed currents with
the result
\begin{equation}\label{naiveN}
    B_{K}(N_c\rightarrow\infty)=\frac{3}{4}\qquad
    \mathrm{and}\qquad C_{\Delta S=2}(\mu)(N_c\rightarrow\infty)=1
    \ .
\end{equation}
Including the next-to-leading order in the $1/N_c$ expansion one
obtains\begin{equation}\label{BK}
    \hat{B}_K= \frac{3}{4} \ C_{\Delta S=2}(\mu)\ g_{\Delta
    S=2}(\mu)\ ,
\end{equation}
where $g_{\Delta S=2}(\mu)$ is given by Eqs. (\ref{g}) and $C_{\Delta S=2}(\mu)$ is given
by Eq. (\ref{wilson}). We emphasize that the function $\mathcal{W}^{LRLR}(Q^2)$ in Eq.
(\ref{ward}) is of $\mathcal{O}(N_c)$ and, therefore, it is given by tree-level diagrams
with infinitely narrow resonances. After the redefinition (\ref{def}), the function
$W(z)$ is of $\mathcal{O}(N_c^0)$  and, recalling that $F_0^2 \sim \mathcal{O}(N_c)$,
$g_{\Delta S=2}(\mu)$ is indeed of the form $\sim 1- \mathcal{O}(1/N_c)$, as one would
expect.

In a combined large-$N_c$ and quark mass ($m_q$) expansion, the result in Eq. (\ref{BK})
is only corrected by terms of $\mathcal{O}(m_q/N_c)$. Furthermore, it is also interesting
to emphasize that if $F_0$ is used instead of $F_K$ in the definition of $\hat{B}_K$ in
Eq. (\ref{definition}), then the above expression (\ref{BK}) is also what is being
computed on the lattice as the $\hat{B}_K$ parameter extrapolated to the chiral limit.
This fact is very convenient as it allows for a comparison between the present
large-$N_c$ method and the lattice results in this limit.

\begin{figure}
\renewcommand{\captionfont}{\small \it}
\renewcommand{\captionlabelfont}{\small \it}
\centering
\includegraphics[width=1.0in]{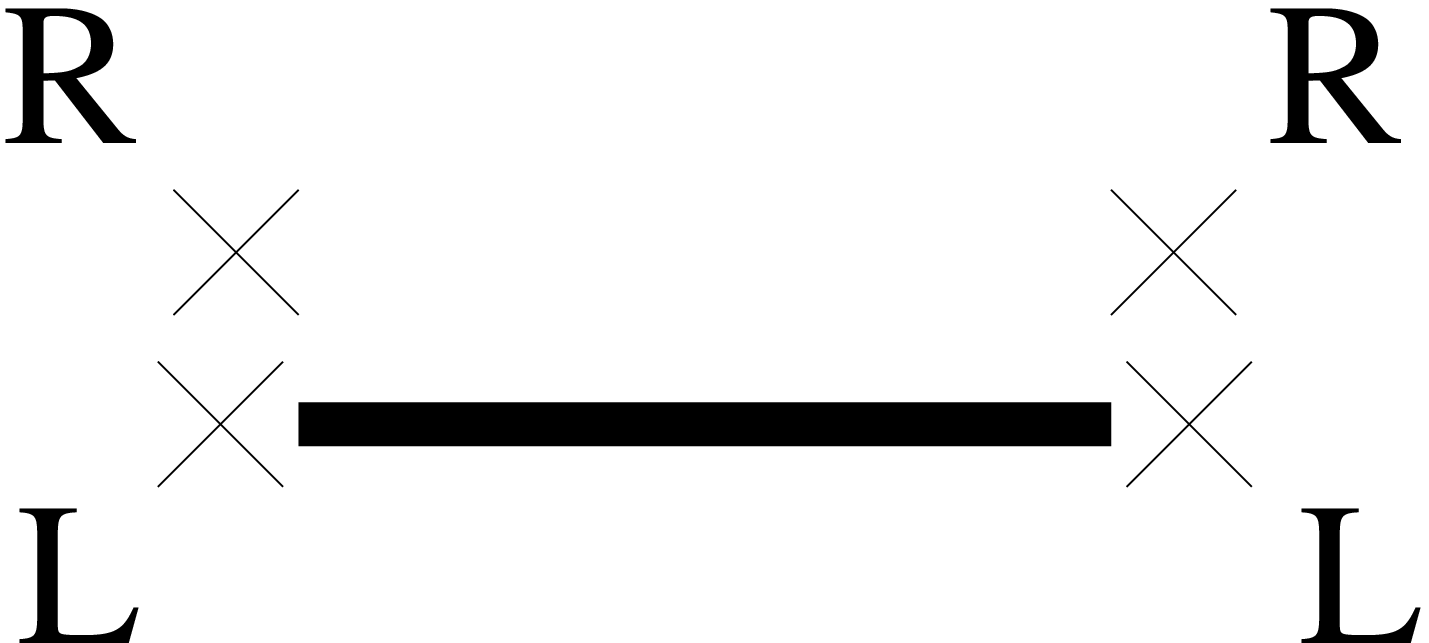}
\hspace{1.in}
\includegraphics[width=1.0in]{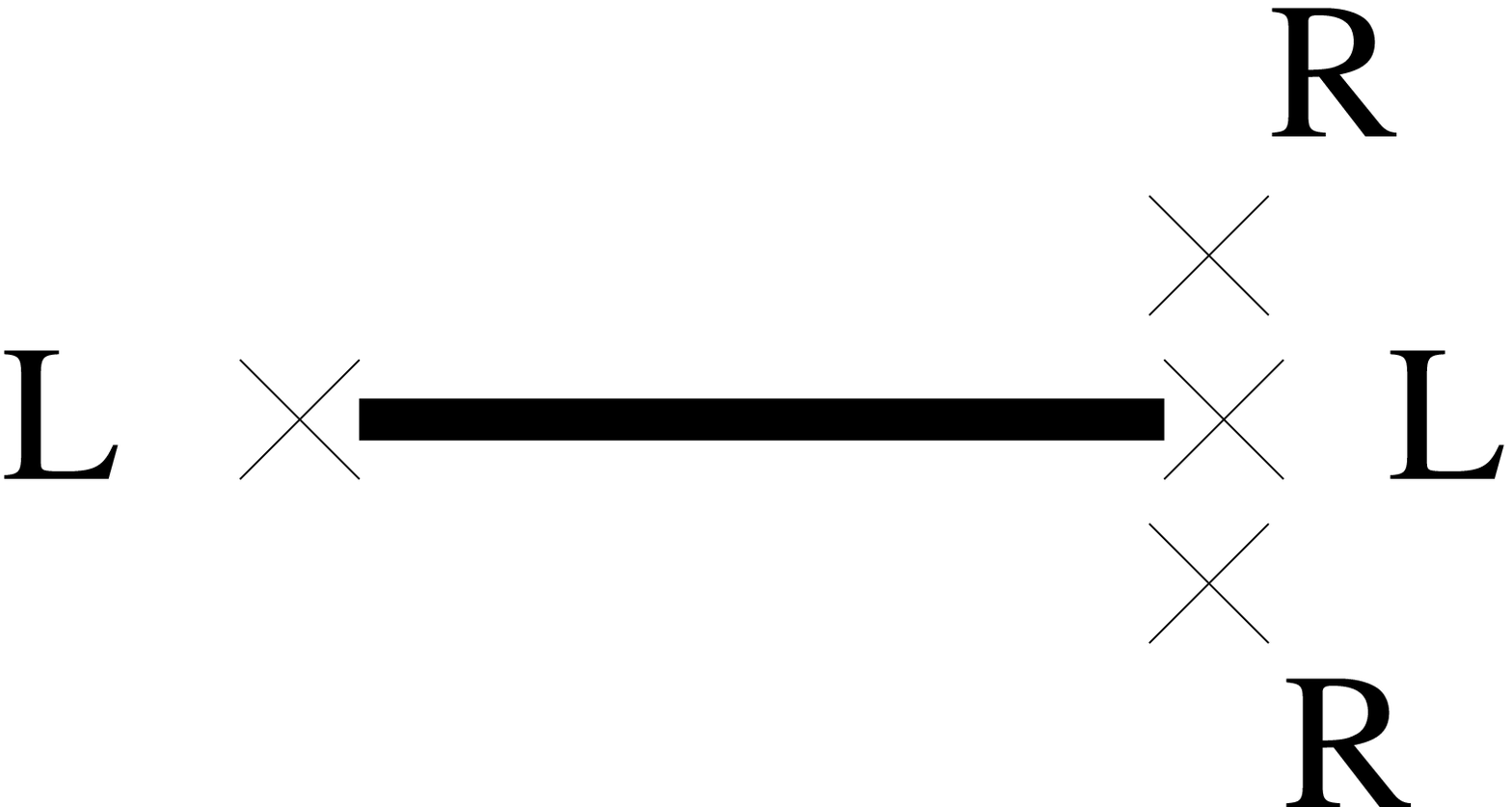}
\hspace{1.in}
\includegraphics[width=1.0in]{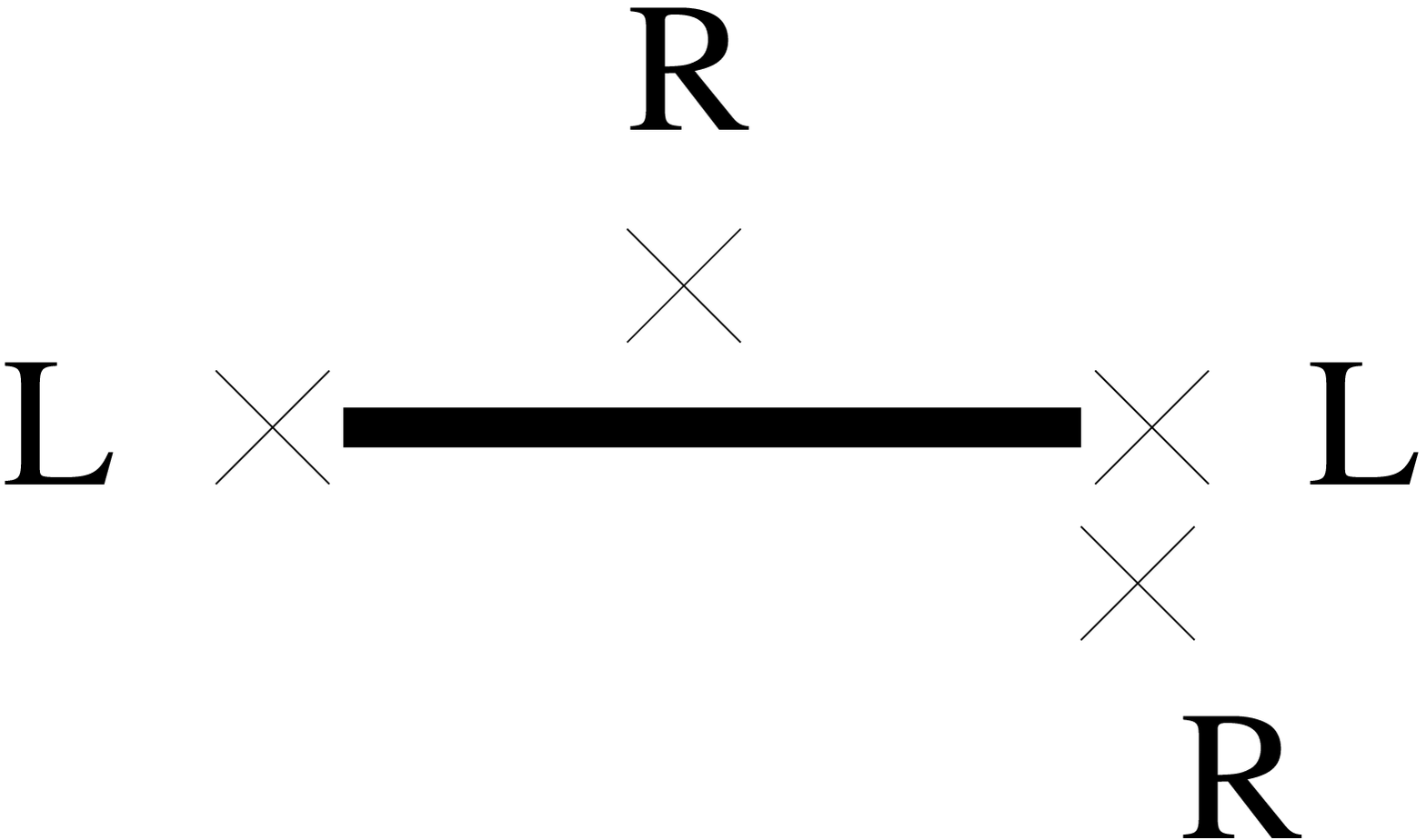}
\hspace{0.5in}
\includegraphics[width=1.0in]{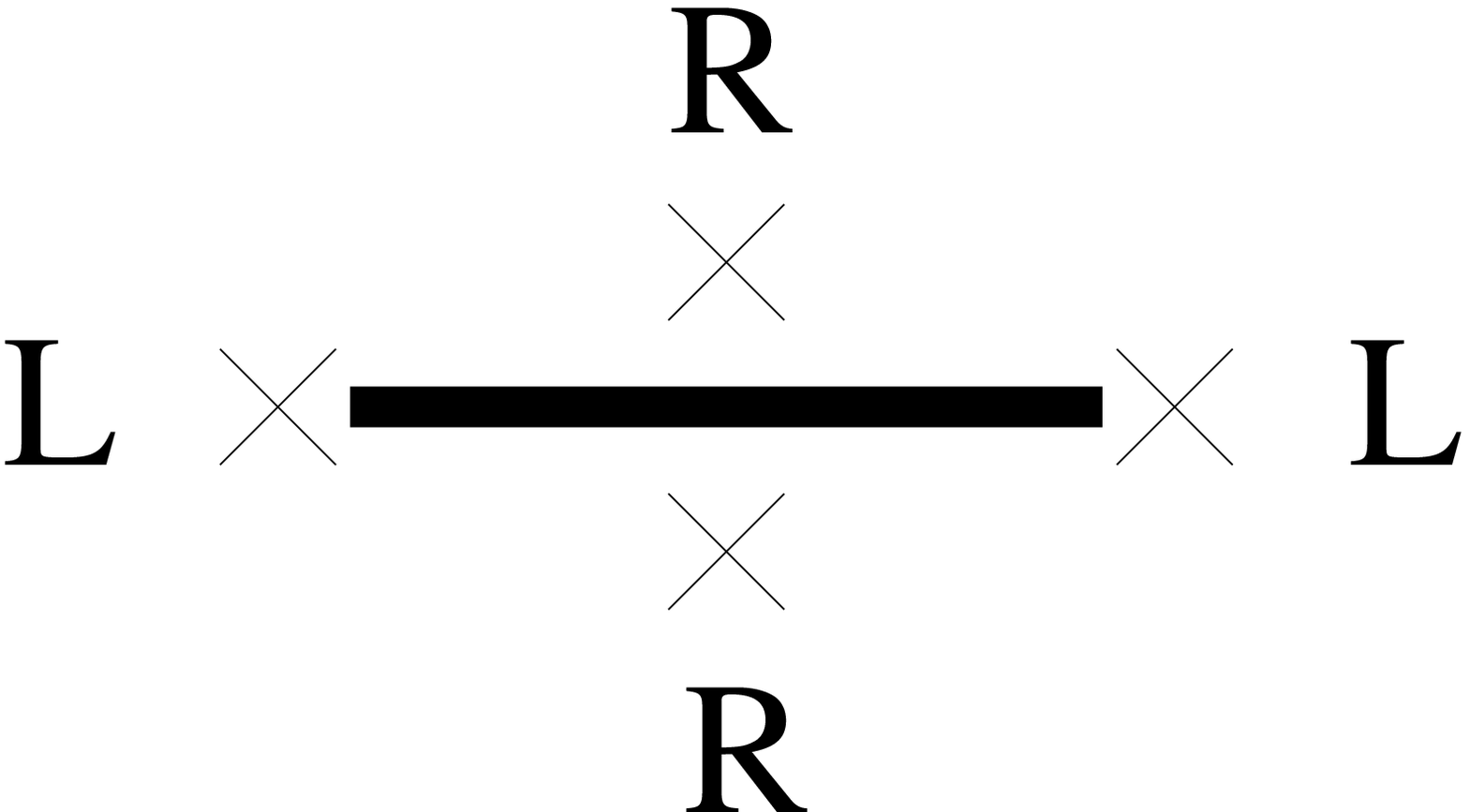}
\hspace{0.5in}
\includegraphics[width=1.0in]{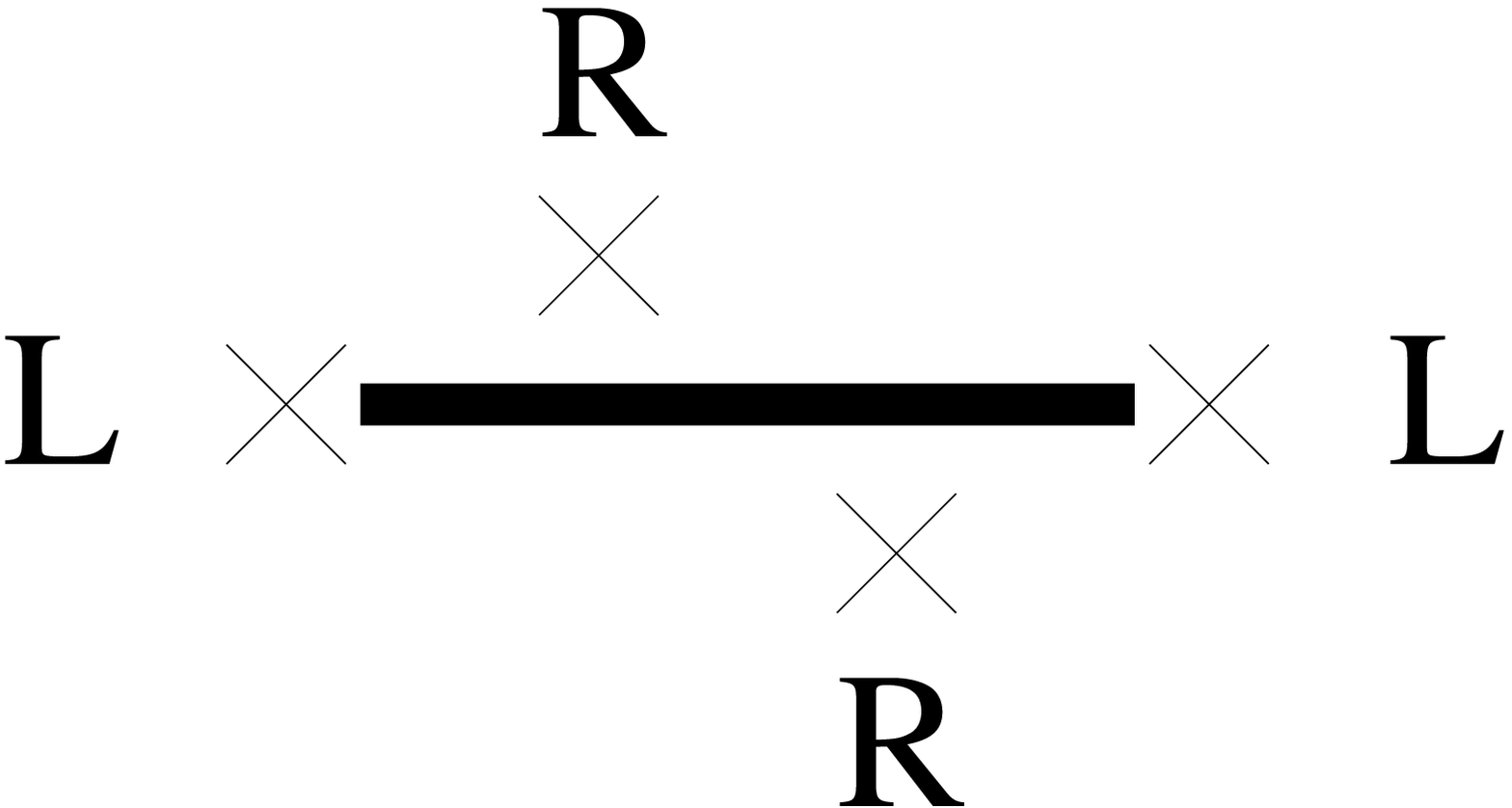}
\caption{{Resonance diagrams contributing to $W(z)$.}}\label{fig:resonance}
\end{figure}

Figure \ref{fig:resonance} shows the different topologies of the resonance diagrams
contributing to $W(z)$ in Eqs. (\ref{Green}-\ref{g}) in the large-$N_c$ limit. Use of the
Mittag-Leffler theorem for meromorphic functions\cite{Mittag} allows one to write $W(z)$
as

\begin{equation} \label{Large}
W(z)= \sum_{i=1}^{\mathcal{N}}\left(\frac{\mathcal{A}_i}{(z+\rho_i)}+
\frac{\mathcal{B}_i}{(z+\rho_i)^2}+\frac{\mathcal{C}_i}{(z+\rho_i)^3}\right)
\end{equation}
where $\rho_i= m_i^2/\mu_{had}^2$, with $m_i$ the mass of the resonance $i$; and
$\mathcal{A}_{i}, \mathcal{B}_{i}, \mathcal{C}_{i}$ are constants. In the large-$N_c$
limit, the sum is in principle extended to an infinite number of resonances, i.e.
$\mathcal{N}\rightarrow \infty$. On the other hand, as mentioned above, the approximation
$W_{HA}(z)$ is defined by restricting the sum in Eq. (\ref{Large}) to a finite
$\mathcal{N}$. In particular, the work of Ref. \cite{PdeR} restricted $\mathcal{N}$ to
just one vector resonance, whose residues where matched onto the leading and
next-to-leading terms in the chiral expansion but only to the leading (nontrivial) term
in the operator product expansion of $W(z)$. This way it was obtained that
$\hat{B}_K=0.38\pm 0.11$; with the error being an estimate of uncalculated terms,
including $1/N_c$ corrections. The purpose of the present work is to compute the
next-to-leading term in the operator product expansion  of $W(z)$ at large-$z$, and its
impact on the previous value for $\hat{B}_K$.

\section{Dimension 8 operators in the weak OPE}
\begin{figure}
\renewcommand{\captionfont}{\small \it}
\renewcommand{\captionlabelfont}{\small \it}
\centering
\includegraphics[width=1.0in]{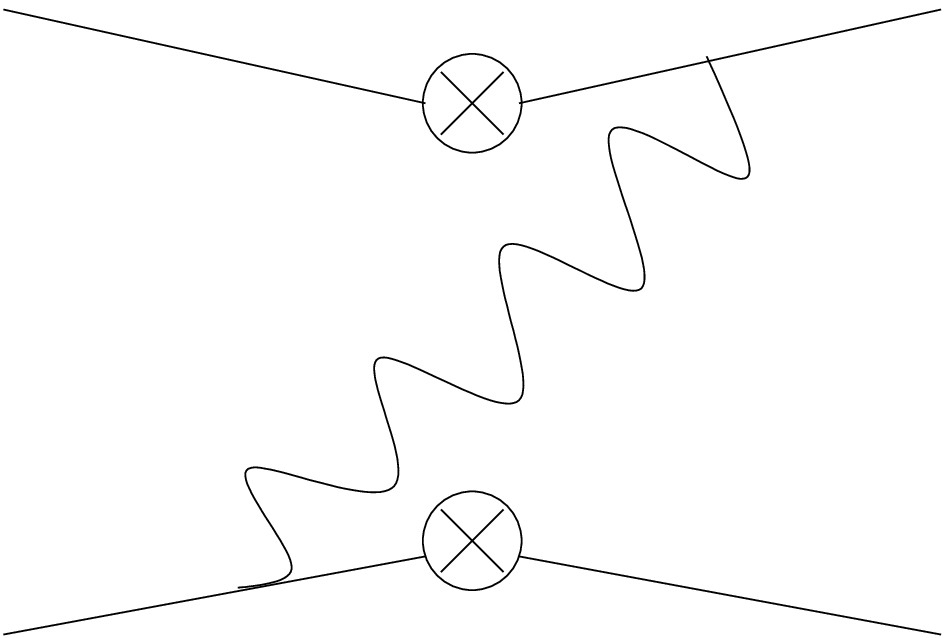}
\hspace{0.5in}
\includegraphics[width=1.0in]{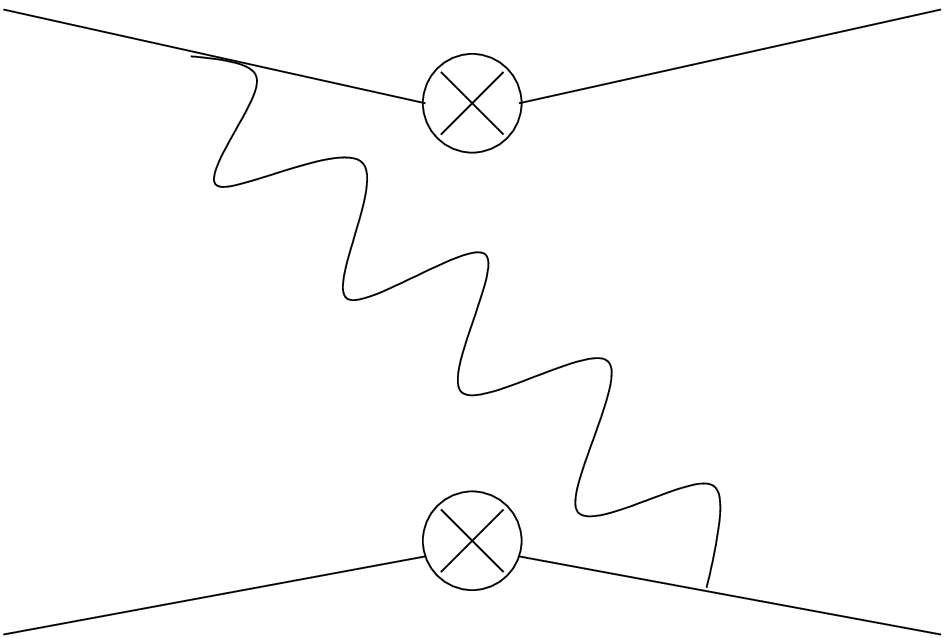}
\hspace{0.5in}
\includegraphics[width=1.0in]{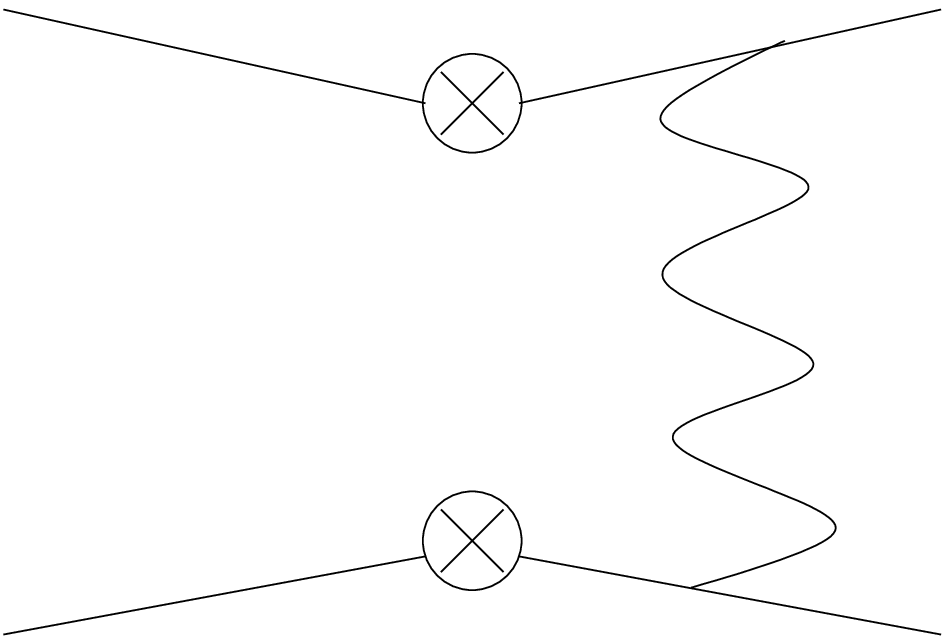}
\hspace{0.5in}
\includegraphics[width=1.0in]{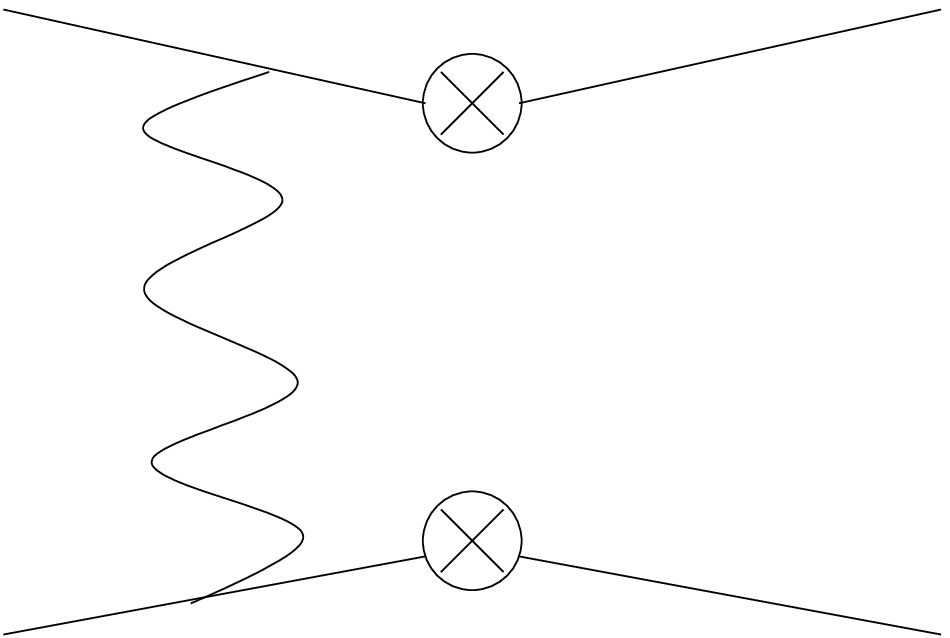}
\caption{{Diagrams contributing to the dimension 6 terms in
    the OPE in Eq. (\ref{ope}).}}\label{fig:op}
\end{figure}

Let us consider the Green's function (\ref{Green}). The large-$Q^2$ behavior of this
function is governed by the Operator Product Expansion
\begin{equation}\label{ope}
\int d^{4}x\  e^{iqx}\
T\left\{L^{\overline{s}d}_{\mu}(x)L_{\overline{s}d}^{\mu}(0)\right\} = \sum_{i}
c_i^{(6)}(Q^2) {\cal{O}}_i^{(6)}+ \sum_{i} c_i^{(8)}(Q^2) {\cal{O}}_i^{(8)}+ ...\ ,
\end{equation}
where the first (second)  sum runs over a set of dimension six (eight) local operators to
be discussed in the following. In the case of dimension-six operators this set is
actually limited to just one, to wit
\begin{equation}\label{dimsix}
    {\cal{O}}^{(6)}={\bar{s}}_L\gamma^{\mu}d_L(0) {\bar{s}}_L\gamma_{\mu}d_L(0)\ ,
\end{equation}
with the Wilson coefficient given, to lowest order in $\alpha_{s}=g_s^2/4\pi$, by
\begin{equation}\label{coeffsix}
c^{(6)}(Q^2)= i \frac{12\pi\alpha_s}{Q^4}.
\end{equation}
The relevant diagrams are listed in Fig. \ref{fig:op}.

The determination of the dimension-eight operators is more involved. We used the
Schwinger's operator method and obtained (after use of the equations of motion)
\begin{eqnarray}\label{operators}
{\cal{O}}_1^{(8)}&=&\bar{s}\overleftarrow{{\cal{D}}_{\mu}}
\overleftarrow{{\cal{D}}^{\mu}}\Gamma^{a}_{\nu}d(0)\ \bar{s}\Gamma^{\nu}_a d(0) +
\bar{s}\Gamma^{a}_{\nu}{\cal{D}}_{\mu}{\cal{D}}^{\mu}d(0)\ \bar{s}\Gamma^{\nu}_a d(0) +
\nonumber  \\
&& \bar{s}\Gamma^{\nu}_a d(0)\ \bar{s}\overleftarrow{{\cal{D}}_{\mu}}
\overleftarrow{{\cal{D}}_{\mu}}\Gamma^{a}_{\nu}d(0) + \bar{s}\Gamma^{\nu}_a
d(0)\ \bar{s}\Gamma^{a}_{\nu}{\cal{D}}_{\mu}{\cal{D}}^{\mu}d(0)\nonumber \ ,\\
{\cal{O}}_2^{(8)}&=&\bar{s}\Gamma^{\nu}_a{\cal{D}}_{\mu} d(0)\
\bar{s}\Gamma^{a}_{\nu}{\cal{D}}^{\mu}d(0) + \bar{s}
\overleftarrow{{\cal{D}}_{\mu}}\Gamma^{a}_{\nu}d(0) \
\bar{s}\overleftarrow{{\cal{D}}^{\mu}}\Gamma^{\nu}_a d(0)
 \nonumber \ ,\\
{\cal{O}}_3^{(8)}&=&\bar{s}\overleftarrow{{\cal{D}}_{\mu}}\Gamma^{\nu}_a d(0)\
\bar{s}\Gamma^{a}_{\nu}{\cal{D}}^{\mu}d(0) + \bar{s}\Gamma^{a}_{\nu}{\cal{D}}_{\mu}d(0)\
{\bar{s}}\overleftarrow{{\cal{D}}^{\mu}}\Gamma_{a}^{\nu}d(0)\nonumber \ ,\\
{\cal{O}}_4^{(8)}&=&\bar{s}\overleftarrow{{\cal{D}}_{\mu}}\Gamma^{\nu}_a d(0)\
\bar{s}\overleftarrow{{\cal{D}}_{\nu}}\Gamma_{a}^{\mu}d(0) +
\bar{s}\Gamma^{a}_{\nu}{\cal{D}}_{\mu}d(0)\ {\bar{s}}\Gamma_{a}^{\mu}{\cal{D}}^{\nu}d(0)\nonumber \ ,\\
{\cal{O}}_5^{(8)}&=&\bar{s}\overleftarrow{{\cal{D}}_{\mu}}\Gamma^{\nu}_a d(0)\
\bar{s}\Gamma_{a}^{\mu}{\cal{D}}_{\nu}d(0) + \bar{s}\Gamma^{a}_{\nu}{\cal{D}}_{\mu}d(0)\
{\bar{s}}\overleftarrow{{\cal{D}}^{\nu}}\Gamma_{a}^{\mu}d(0)\nonumber \ ,\\
{\cal{O}}_6^{(8)}&=&g_s\ \tilde{G}^{a}_{\mu\nu}(0) \ \Big\{\bar{s}\Gamma^{\mu}_a d(0) \
\bar{s}\Gamma^{\nu} d(0) - \bar{s}\Gamma^{\mu} d(0)\ \bar{s}\Gamma^{\nu}_a d(0)\Big\}
\end{eqnarray}
where the following conventions were adopted:
\begin{eqnarray}
\Gamma^{\mu}_a& = &\frac{\lambda_a}{2}\ \gamma^{\mu}\ \frac{1-\gamma_5}{2}\qquad
,\qquad {\cal{D}}_{\mu}  = \partial_{\mu}-ig_sA_{\mu}\qquad , \qquad
A_{\mu} = \frac{A_{\mu}^{a} \lambda_{a}}{2} \nonumber\\
G_{\mu\nu}& = &\partial_{\mu}A_{\nu}-\partial_{\nu}A_{\mu}-ig_s[A_{\mu},A_{\nu}]\quad ,
\quad {\tilde{G}}_{\mu\nu}=\frac{1}{2}G^{\rho\sigma}\epsilon_{\rho\sigma\mu\nu}\ ,\
\mathrm{with} \quad \epsilon^{0123}= +1\ .
\end{eqnarray}
The corresponding Wilson coefficients are
\begin{equation}\label{coeffs}
    c_i^{(8)}= i \frac{4\pi\alpha_s}{Q^6}\ \eta_i^{(8)}\ , \nonumber \\
\end{equation}
where
\begin{eqnarray}\label{coeffeight}
\eta_1^{(8)}=\frac{5}{3}\qquad , \qquad \eta_2^{(8)}&=&\frac{22}{3}\qquad , \qquad
\eta_3^{(8)}=\frac{8}{3} \nonumber\\
\eta_4^{(8)}=\frac{18}{3}\qquad ,\qquad \eta_5^{(8)}&=&\frac{16}{3}\qquad ,\qquad
\eta_6^{(8)}=\frac{1}{N_c}\quad.
\end{eqnarray}
This result agrees with the calculation in Ref. \cite{CDG} if one takes into account
that, in the present case, the specific flavor structure $(\overline{s}d)^2$ leads to
some simplifications. In particular, it allows one to rewrite $\mathcal{Q}_4$ in Ref.
\cite{CDG} in terms of the $\mathcal{O}_{4,5}$ operators in Eq. (\ref{operators}).

Inserting the operators (\ref{dimsix},\ref{operators}) in Eqs.
(\ref{ope},\ref{Green}-\ref{g}) one can compute the expansion in powers of $1/z$ of the
function $W(z)$ in Eq. (\ref{g}). The terms of $\mathcal{O}(z^{-1})$ were already
computed in Ref. \cite{PdeR}. In doing the calculation of the next-to-leading terms, of
$\mathcal{O}(z^{-2})$, we notice that several further simplifications take place.
Firstly, since the contribution is proportional to $\alpha_{s}$ -- see Eq.
(\ref{coeffsix},\ref{coeffs})-- , the large-$N_c$ limit allows one to use factorization.
Secondly, the contribution from $\mathcal{O}_{6}$ is $1/N_c$ suppressed due to the value
of its Wilson coefficient,  Eq. (\ref{coeffeight}). Thirdly, one is only interested in
the leading term in the $l\rightarrow 0$ limit of the Green's function (\ref{Green}).
Since, after Fierzing, one ends up with matrix elements of the form
\begin{equation}\label{nada}
    <0|\bar{s}_{L}\Gamma_{\nu}{\cal{D}}_{\mu}d_{L}(0)|K(l)>\ \sim a\  g_{\mu\nu} +
    b\ l_{\mu} l_{\nu}\ ,
\end{equation}
by contracting with $g^{\mu\nu}$  one immediately concludes that $a\sim \mathcal{O}(l^2)$
and, consequently, $\mathcal{O}_{2,3,4,5}$ can be neglected since they yield
contributions to Eq. (\ref{ward}) with an extra power of $l^2$. Therefore, the term of
$\mathcal{O}(z^{-2})$ in  the large-$z$ expansion of $W(z)$ in Eq. (\ref{g}) is governed
solely by $\mathcal{O}_{1}$. Furthermore, using $D^{2}=D\!\!\!\!/\ ^2+ \frac{g_s}{2}
\sigma_{\mu\nu} G^{\mu\nu}$ in the expression for $\mathcal{O}_1$, and the previous
considerations, lead to the conclusion that the final result can be given in terms of a
\emph{single} matrix element\footnote{This is related to the fact that
$\overline{s}\gamma^{\mu}\widetilde{G}_{\mu\nu}d$ is the only nontrivial dimension-five
operator in the chiral limit with these quantum numbers \cite{NSVZ1}.}, namely
\begin{equation} \label{matrixelement}
<0| g_s \bar{s}_L\tilde{G}_{\mu\nu}^a \lambda_a \gamma^{\mu}d_L |K(l)
>=  - i \sqrt{2}F_0\ \delta_{K}^2\ l_{\mu}\ ,
\end{equation}
where $\delta_{K}^2$ is a parameter to be determined in the next section. In conclusion,
the OPE of $W(z)$ in Eq. (\ref{g}) can be written in terms of this parameter
$\delta_{K}^2$ as
\begin{equation}\label{newope}
W^{OPE}(z)=\frac{24\pi\alpha_sF_0^2}{\mu_{had}^2}\ \frac{1}{z}
    \left[1+\frac{\epsilon}{12}(5+\kappa)+
    \frac{10}{9} \frac{\delta_{K}^2}{\mu_{had}^2}\ \frac{1}{z}\right] +
    \mathcal{O}\left(\frac{1}{z^3}\right)
    \ .
\end{equation}
Since the term proportional to $\delta_{K}^2$ yields an ultraviolet  convergent integral
in Eq. (\ref{g}), it can be computed in $4$ dimensions. This is why there is no
$\epsilon$ dependence in Eq. (\ref{newope}) accompanying this term.

\section{Determination of $\delta_{K}^{2}$}

The mixed quark-gluon matrix element (\ref{matrixelement}) was determined in the
literature in {\cite{NSVZ1, Nar}}. Both approaches relied on  an analysis of Borel Sum
Rules applied to two-point functions with nonvanishing contribution from the perturbative
continuum. Although their final result for $\delta_{K}^2$ happened to be in agreement,
there were some criticisms raised by \cite{Nar} on the analysis of \cite{NSVZ1}, in
particular on the choice for the onset of this perturbative continuum, $s_0$. This partly
motivated us to do a reanalysis of the matrix element (\ref{matrixelement}).

A particularly simple way to bypass the need for choosing a value for $s_0$ consists in
looking at Green's functions which are order parameters of spontaneous chiral symmetry
breaking since in these functions the perturbative continuum vanishes by construction.
This procedure has proved fruitful in several contexts {\cite{EdeR}} and will be applied
to the present case as well.

Consider the Green's function
\begin{equation} \label{GreenLR}
\widetilde{\Pi}_{\mu\nu}^{LR}(Q^2)=i\int\ d^4x\ e^{iq·x}
<0\,|T\left\{\frac{g_s}{2}\bar{s}_L\tilde{G}_{\alpha\mu}\gamma^{\alpha}d_L(x)\
\bar{d}_R\gamma_{\nu}s_R(0)\right\}|0> \ .
\end{equation}
Lorentz invariance guarantees that, in the chiral limit, it has the following tensorial
structure
\begin{equation}\label{pitilde}
\widetilde{\Pi}_{\mu\nu}^{LR}(q)=(q_{\mu}q_{\nu}-g_{\mu\nu}q^2)\
\widetilde{\Pi}^{LR}(q^2)\ .
\end{equation}
The large-$q^2$ fall-off of the function (\ref{pitilde}) can be straightforwardly
computed to be
\begin{equation} \label{LROPE}
\widetilde{\Pi}^{LR}(q^2)= - \frac{2\pi}{9}\frac{\alpha_{s}<\bar{\psi}\psi>^2}{Q^4}\quad
, \quad Q^2\equiv -q^2\ ,
\end{equation}
where factorization of the four-quark condensate in the large-$N_c$ limit has been used.
Since $\widetilde{\Pi}_{LR}$ obeys the unsubtracted dispersion relation
\begin{equation} \label{dispersion}
\widetilde{\Pi}^{LR}(Q^2)=\int_0^{\infty}dt\frac{1}{\pi}\frac{Im\,\widetilde{\Pi}^{LR}(t)}{t+Q^2}\
,
\end{equation}
we can consider as the minimal hadronic approximation the following spectral function
\begin{equation} \label{spectral}
\frac{1}{\pi}Im\,\widetilde{\Pi}^{LR}(t)=-\frac{F_0^2\delta_{K}^2}{4}\delta(t)+
\frac{f_V^2\delta_V^2}{8}\delta(t-m_V^2)
\end{equation}
in which one introduces, besides the Goldstone boson, a vector resonance. Inserting
(\ref{dispersion}) into (\ref{spectral}) and expanding for large $Q^2$ one obtains, upon
comparison with the OPE of (\ref{LROPE}), the two Weinberg-like sum rules
\begin{eqnarray}
\frac{F_0^2\delta_{K}^2}{4}&=&\frac{f_V^2\delta_V^2}{8}\nonumber\\
\frac{f_V^2\delta_V^2}{8}m_V^2&=&\frac{2\pi}{9}\ \alpha_s <\bar{\psi}\psi>^2\ ,
\end{eqnarray}
from which the unknown $\delta_V^2$ and $\delta_K^2$ parameters are readily determined to
be
\begin{eqnarray}\label{deltas}
\delta_V^2&=&\frac{16\pi}{9}\ \frac{\alpha_s<\bar{\psi}\psi>^2}{f_V^2m_V^2}\ ,\nonumber\\
\delta_{K}^2&=&\frac{8\pi}{9}\ \frac{\alpha_s <\bar{\psi}\psi>^2}{F_0^2 m_V^2}\ .
\end{eqnarray}
Using $F_0 \simeq 0.087$ GeV, $f_V \simeq 0.15$, $m_V \simeq 0.77$ GeV, $\alpha_s(2\
\mathrm{GeV})\simeq 1/3$ and $\langle\bar{\psi}\psi\rangle(2\ \mathrm{GeV})= - (280 \pm
30\ \mathrm{MeV})^3$  one obtains\footnote{We have added generous error bars in the quark
condensate to include the present spread of values in this quantity\cite{condensate}.
This error is the dominant one.}
\begin{equation}\label{numbers}
\delta_K^2=0.12 \pm 0.07\ \mathrm{GeV}^2\qquad ,\qquad \delta_V^2= 0.06 \pm 0.04\
\mathrm{GeV}^4\ .
\end{equation}
Strictly speaking, both parameters $\delta_{K}^2$ and $\delta_{V}^2$ depend on the
renormalization scale $\mu$. This dependence, however, is very small. As a matter of
fact, according to Ref. \cite{shuryak} the combination
\begin{equation}\label{mudependence}
\alpha_s(\mu)^{-\frac{4}{11}}\ g_s\bar{s}_L\gamma^{\mu}\tilde{G}_{\mu\nu}d_L
\end{equation}
is renormalization group invariant. This means that the $\mu$ dependence of the
parameters $\delta_{K,V}^2$ in Eq. (\ref{deltas}) yields a variation of $10\%$ if $\mu$
is varied in the range $1\leq\mu\leq 2\ \mathrm GeV$.

\begin{figure}\label{duality}
\renewcommand{\captionfont}{\small \it}
\renewcommand{\captionlabelfont}{\small \it}
\centering
\psfrag{t}{\Large {$\mathbf{\tau}$}}
\includegraphics[width=3 in]{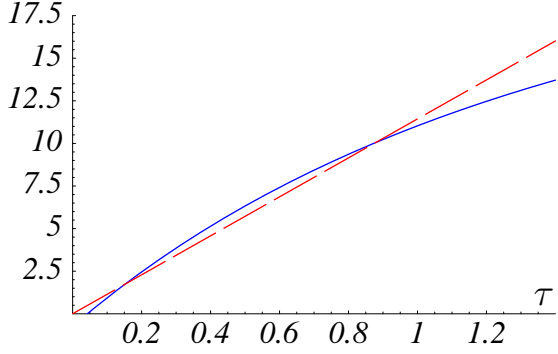}
\caption{Plot of the right and lefthand side of Eq. (\ref{BSR}) (in
arbitrary units), corresponding to the dashed and solid curve,
respectively.}\label{plot1}
\end{figure}

Apart from the dependence on the quark condensate, in order to check how much the result
(\ref{numbers}) depends on the two-particle decomposition of the spectral function in Eq.
(\ref{spectral}), we can now repeat the analysis by introducing a third state. We take
this state to be the first axial resonance. In this case, the spectral function reads
\begin{equation} \label{spectr}
\frac{1}{\pi}Im\,\widetilde{\Pi}^{LR}(t)= - \frac{F_0^2\delta_{K}^2}{4}\ \delta(t) +
\frac{f_V^2\delta_V^2}{8} \delta(t-m_V^2) + \frac{f_A^2\delta_A^2}{8} \delta(t-m_A^2) \ .
\end{equation}
One can now apply the Borel transform to (\ref{spectr}) and the OPE of (\ref{LROPE}) to
get the relation
\begin{equation}  \label{BSR}
-\frac{F_0^2\delta_{K}^2}{4}+\frac{f_A^2\delta_A^2}{8} e^{-m_A^2\tau}+
\frac{f_V^2\delta_V^2}{8} e^{-m_V^2\tau}=-\frac{2\pi}{9} \alpha_{s}<\bar{\psi}\psi>^2\tau
\ ,
\end{equation}
where $\tau$ is the Borel parameter. The equality above is fulfilled within a rather wide
window of duality, namely  $0\lesssim\tau\lesssim 1$, as can be seen in figure
(\ref{duality}). Using $f_A= 0.08$ and $m_A= 1.2$ GeV, from this plot one extracts values
for $\delta_{K}^{2}, \delta_{V}^{2}$ which agree with the ones obtained in
(\ref{numbers}). There is, therefore, consistency in the result obtained with the two
methods. For the new parameter, $\delta_{A}^{2}$, one obtains $\sim 0.05\
\mathrm{GeV}^4$. We remark that the value of $\delta_{K}^{2}$ is a bit lower than the
results obtained in refs. \cite{NSVZ1, Nar}. To the best of our knowledge there is no
lattice evaluation of $\delta_K^2$.

\section{Numerical Evaluation of $\widehat{B}_K$ and Conclusions}

In Ref. \cite{PdeR} the correction from dimension-eight operators was not considered and
matching to the OPE was accomplished with just a single vector resonance whose mass was
taken to be the physical $\rho$ mass. In the present case we have a further constraint
given by the new large-$z$ behavior in Eq. (\ref{newope}) caused by the presence of the
parameter $\delta_K^2$. Although one might think that this new constraint could allow one
to determine the mass of the vector resonance, it turns out that this is actually not
possible: the positive value of $\delta_K^2$ would force the $\rho$ mass to be purely
imaginary. Therefore one must consider a further resonance.

 Besides vectors, scalar
particles also contribute to the coupling $L_3$\cite{Sui1, Sui2}. The dependence of the
function $W(z)$ on $L_3$ at low-$z$ found in \cite{PdeR} --see Eq. (\ref{lowz})--
suggests, therefore, a scalar particle as the new resonance in the sum (\ref{Large}).
Furthermore, a scalar particle appears as a single pole in the function $W(z)$ in Eq.
(\ref{Large}) as a consequence of the quantum numbers being exchanged in the Green's
function (\ref{Green}) (see Fig. 1). Interestingly, one scalar residue is all one needs
to balance the new constraint from the OPE encoded in $\delta_K^2$ through Eq.
(\ref{newope}). As to the scalar mass, which remains undetermined, we allow for the
generous variation $m_s= 900 \pm 400$ MeV.

Gathering all the pieces, one has as the interpolating function
\begin{equation}\label{interpolator}
W_{HA}(z;\{S,V\})=\frac{\mathcal{A}_V}{(z+\rho_V)}+\frac{\mathcal{B}_V}{(z+\rho_V)^2}+
\frac{\mathcal{C}_V}{(z+\rho_V)^3}+\frac{\mathcal{A}_S}{(z+\rho_S)}\ ,
\end{equation}
to be matched onto
\begin{equation}\label{lowz}
 W^{\chi PT}(z)= 6 - 24\
\frac{\mu_{had}^2}{F_0^2}\ \Big(2 L_1+5 L_2+L_3+L_9\Big)\ z + {\cal{O}} \left(z^2\right)
\end{equation}
at low energies, and
\begin{equation}\label{highz}
 W^{OPE}(z)=\frac{24\pi\alpha_sF_0^2}{\mu_{had}^2}\ \frac{1}{z}\
    \left[1+\frac{\epsilon}{12}(5+\kappa)+
    \frac{10}{9}\ \frac{\delta_{K}^2}{\mu_{had}^2}\ \frac{1}{z}\right]
     + \mathcal{O}\left(\frac{1}{z^3}\right)
\end{equation}
at high energies. This results in the following 4 constraints:
\begin{eqnarray}\label{finalmatching}
{\mathcal{A}_V}+ \mathcal{A}_S &=&\frac{24\pi\alpha_s
  F_0^2}{\mu_{had}^2}\left[1+\frac{\epsilon}{12}(5+\kappa)\right]\nonumber\\
{\mathcal{B}_V}- {\mathcal{A}_V}\rho_V- {\mathcal{A}_S}\rho_S&=&\frac{24\pi\alpha_s
F_0^2}{\mu_{had}^4}
\left(\frac{10}{9}\delta_K^2\right)\nonumber\\
\frac{\mathcal{A}_V}{\rho_V}+\frac{\mathcal{B}_V}{\rho_V^2}+\frac{\mathcal{C}_V}{\rho_V^3}+
\frac{\mathcal{A}_S}{\rho_S}&=&6\nonumber\\
\frac{\mathcal{A}_V}{\rho_V^2}+2\frac{\mathcal{B}_V}{\rho_V^3}+
3\frac{\mathcal{C}_V}{\rho_V^4}
+\frac{\mathcal{A}_S}{\rho_S^2}&=&24\frac{\mu_{had}^2}{F_0^2}\left(2L_1+5L_2+L_3+L_9\right)\
,
\end{eqnarray}
for the 4 unknowns $\mathcal{A}_S, \mathcal{A}_V, \mathcal{B}_V$ and $\mathcal{C}_V$. The
combination $\left(2L_1+5L_2+L_3+L_9\right)$ in the last equation equals $11.2 \times
10^{-3}$ if its experimental value evaluated at the $\rho$ mass scale is used \cite{Pich}
\footnote{If the results of Ref. \cite{ABT,GP} are used, this
    combination is $8.7\times 10^{-3}$ ($11.7\times 10^{-3}$), respectively. }. This combination
    is renormalization scale dependent, but this
    dependence can be neglected as it is subleading in the $1/N_c$ expansion.
    Numerically, it
    amounts to $30\%$ if $\mu$ is varied in the range $0.5\leq \mu\leq
    1\ \mathrm{GeV}$. The coupling $\alpha_{s}$ is evaluated at the scale $\mu_{had}$,
    which is
chosen in the range $1-2\ \mathrm{GeV}$. All these variations will be included in the
final error.

 Figure 4 shows the low- and high-$z$ behavior of
the function $W(z)$ given by its chiral and operator product expansion (dashed curves)
together with the interpolator $W_{HA}(z)$ in Eq. (\ref{interpolator}) (solid curve).

\begin{figure}
\renewcommand{\captionfont}{\small \it}
\renewcommand{\captionlabelfont}{\small \it}
\centering
\psfrag{A}{\Huge $\frac{W(z)}{W(0)}$}
\includegraphics[width=4.5 in]{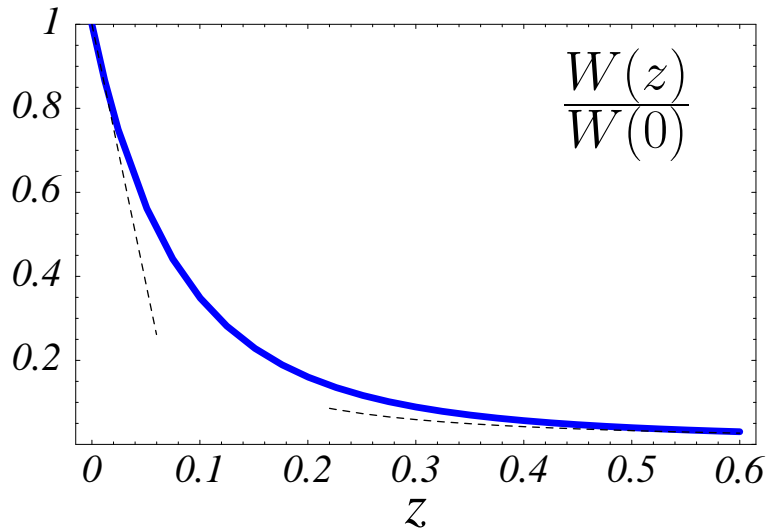}
\caption{Plot of the large- and small-$z$ expansions of
the function $W(z)$ in Eq. (\ref{g}) as given by the OPE and Chiral Theory, respectively
(dashed curves). The solid curve corresponds to the interpolating function $W_{HA}(z)$
obtained in Eq. (\ref{interpolator}). }\label{plot1}
\end{figure}
The main effect from the dimension-eight operators is to push the large-$z$ tail of the
function $W(z)$ slightly upwards, as a consequence of $\delta_K^2$ being positive. This
implies a larger value of the integral in Eq. (\ref{g}), i.e. a larger area under the
curve and, consequently, a smaller value for $\widehat{B}_K$. One can  get a rough idea
about the size of the contribution from the dimension-eight operators by noticing that
the $\delta_K^2$ in Eq. (\ref{highz}) generates a correction to the coupling $g_{\Delta
S=2}$ in Eq. (\ref{g}) given by
\begin{eqnarray}\label{estimate}
  \delta g_{\Delta S=2} &\sim& - \ \frac{\mu_{had}^2}{32 \pi^2 F_0^2}\ \
  \ \frac{24 \pi \alpha_{s}(\mu_{had})F_0^2}{\mu_{had}^2}\ \
  \ \frac{10}{9}\ \frac{\delta_{K}^2}{\mu_{had}^2}\
  \int_{\sim \frac{\mu_{ope}^2}{\mu_{had}^2}}^{\infty}\ \frac{dz}{z^2} \nonumber \ , \\
   & \sim & - \ \frac{15}{18}\frac{\alpha_{s}(\mu_{had})}{\pi}
   \ \frac{\delta_{K}^2}{\mu_{ope}^2}\quad ,
\end{eqnarray}
where $\mu_{ope}$ is the scale above which the large-$z$ expansion (i.e. the OPE) starts
making sense. Numerically, one obtains $|\delta g| \lesssim 0.03$ when $\mu_{ope}\sim 1$
GeV and the value for $\delta_{K}^2$ in Eq. (\ref{numbers}) are used.

 To get a more accurate result one has to use
 the solution for $\mathcal{A}_S, \mathcal{A}_V, \mathcal{B}_V$ and
 $\mathcal{C}_V$ from
Eqs. (\ref{finalmatching}) to construct the hadronic approximation $W_{HA}$ in Eq.
(\ref{interpolator}). One can then explicitly perform the integral in (\ref{g}) to
obtain, after multiplication by the Wilson coefficient in Eq. (\ref{wilson}) according to
Eq. (\ref{BK}), a cancelation of scale and scheme dependence\cite{PdeR} with the
following result\footnote{The quadratic dependence on $\mu_{had}$ is actually fictitious
as is canceled by that of $\mathcal{A}_{V,S},\mathcal{B}_V$ and $\mathcal{C}_V$; see Eq.
(\ref{finalmatching}).}
\begin{eqnarray}\label{result}
    \widehat{B}_K&=&\left(\frac{1}{\alpha_s(\mu_{had})}\right)^{\frac{3}{11}}\frac{3}{4}
\Biggl[1-\frac{\alpha_s(\mu_{had})}{\pi}\frac{1229}{1936}
+{\cal{O}}\left(\frac{N_c\alpha_s^2(\mu_{had})}{\pi^2}\right)- \nonumber\\&-&
\frac{\mu_{had}^2}{32\pi^2F_0^2}\left(- \mathcal{A}_V \log \rho_V - \mathcal{A}_S \log
\rho_S +
\frac{\mathcal{B}_V}{\rho_V}+\frac{1}{2}\frac{\mathcal{C}_V}{\rho_V^2}\right)\Biggr]\ .
\end{eqnarray}

Using now the values $\alpha_s(2\ \mathrm{GeV})\simeq 0.33$, $F_0=0.087\ \mathrm{GeV}$,
$1\ \mathrm{GeV}\leq \mu_{had}\leq 2\ \mathrm{GeV}$, the masses $m_V= 0.77 \pm 0.03$ GeV,
$m_S= 0.9 \pm 0.4$ GeV and the matrix element $\delta_K^2=0.12\pm 0.07 \ \mathrm{GeV}^2$,
the expression (\ref{result}) yields
\begin{equation}\label{final}
    {\hat{B}}_K= 0.36 \pm 0.15 \ ,
\end{equation}
where the error quoted embraces an estimate of uncalculated $1/N_c$ corrections and the
``noise'' created by the uncertainty in all the parameters, taken one at a time.

We would like to conclude with a comparison with other results. Since the $\Delta S=2$
operator (\ref{bosonization}) sits in the $(27_{L},1_{R})$ representation of flavor
$SU(3)_L\times SU(3)_R$, the result (\ref{final}) translates into a value for the
coupling constant $g_{27}$ which governs $\Delta I=3/2$ processes, such as
$K^+\rightarrow \pi^+ \pi^0$, namely\cite{Donoghue,PichdeR}
\begin{equation}\label{pich}
    g_{27}= \frac{4}{5}{\hat{B}}_K = 0.29 \pm 0.12\ .
\end{equation}
This number is in very good agreement with a recent extraction \cite{bij}  of $g_{27}$
from $K \to 3\pi$ decays which yields, after subtraction of chiral corrections (within
reasonable assumptions), the value $g_{27} \simeq 0.24$.

However, because chiral symmetry is much more difficult to have on the lattice than on
the continuum \cite{panel}, the situation concerning the value of ${\hat{B}}_K$ in
numerical simulations in the chiral limit is not totally clear. It is seen in lattice
data that ${\hat{B}}_K$ dips as quark masses go to zero\cite{lellouch}, but the errors do
not yet allow a sufficiently accurate extraction\footnote{See Ref. \cite{hope} for a new
strategy to try to overcome the difficulties.}. Nevertheless, we remark that two lattice
collaborations have recently obtained ${\hat{B}}_K\simeq 0.3-0.4$ when extrapolated to
the chiral limit\cite{RBC}.

Finally, alhough ${\hat{B}}_K$ in the chiral limit may be an interesting problem per se,
it is clear that Nature is not in this limit and, therefore, that it is of the utmost
importance to compute chiral corrections to the result (\ref{final}). As a matter of
fact, lattice results \cite{lellouch2} and phenomenological analysis \cite{martinelli}
seem to favor a large value for ${\hat{B}}_K$ at the physical kaon mass of the order of
twice the value in (\ref{final}). This would imply that the $\mathcal{O}(m_q/N_c)$
corrections to our result (\ref{final}) are very large; a result which should be
understood by analytic methods. That this is not impossible was shown in the calculation
of Ref. \cite{enjl}. Even though within a model with notable differences with respect to
QCD \cite{cenjl}, the authors of Ref. \cite{enjl} obtained a value of ${\hat{B}}_K$ in
the chiral limit in agreement with our result (\ref{final}) while at the same time
producing a larger ${\hat{B}}_K$ at the physical kaon mass, in agreement with Refs.
\cite{lellouch2,martinelli}.

\section{Acknowledgements}
We would like to thank M. Knecht and E. de Rafael for discussions and a careful reading
of the manuscript. We would also like to thank E. de Rafael for pointing out the
convenience of the Green's function (\ref{GreenLR}) in the determination of $\delta_K^2$.
We would also like to thank the warm hospitality received at the CPT at Marseille. Work
supported by CICYT-FEDER-FPA2002-00748, 2001-SGR00188, TMR EC-Contract No.
HPRN-CT-2002-00311 (EURIDICE) and the France-Spain accion integrada ``Picasso''.

\end{document}